\documentclass[12pt,preprint]{iopart}
\usepackage{color}
\usepackage{graphicx}
\usepackage{hyperref}

\def    \bta          {\mbox{\boldmath$\tau$}}

\begin{document}

\title{Impurity states in graphene with intrinsic spin-orbit interaction}
\author{M Inglot$^1$
\footnote{Corresponding author: ming@prz.edu.pl}
 and V K Dugaev$^{1,2}$}
\address{$^1$ Department of Physics, Rzesz\'ow University of Technology,
Al.~Powsta\'nc\'ow Warszawy 6, 35-959 Rzesz\'ow, Poland}
\address{$^2$ Department of Physics and CFIF, Instituto Superior T\'ecnico,
Technical University of Lisbon, Av.~Rovisco Pais, 1049-001 Lisbon, Portugal}
\ead{ming@prz.edu.pl}
\date{\today }

\begin{abstract}
We consider the problem of electron energy states related to strongly localized potential
of a single impurity in
graphene. Our model simulates the effect of impurity atom substituting the atom
of carbon, on the energy spectrum of electrons
near the Dirac point. We take into account the internal
spin-orbit interaction, which can modify the structure of electron bands at very small
neighborhood of the Dirac point, leading to the energy gap. This makes possible the occurrence
of additional impurity states in the vicinity of the gap.
\end{abstract}

\maketitle

\section{Introduction}

Graphene attracted a lot of attention recently due to very unusual properties
of electron energy spectrum and transport properties, including both the transport of
electrons and phonons \cite{K.S.Novoselov10222004,NovoselovNature05,GeimNatMater07,Katsnelson200720,SchedinNature07}.
The most striking properties of graphene are related to the
energy spectrum near the Dirac points, where this spectrum is linear as a function of momentum,
and the Hamiltonian of free electrons can be described by the relativistic
two-dimensional Dirac model \cite{RevModPhys.81.109}.

Naturally, the impurities and defects can strongly affect
the energy spectrum of graphene. Especially important is the
effect of impurities on the spectrum near the Dirac point.
The impurity states and the corresponding variation of
the electron density of states have been already discussed in several papers
\cite{PhysRevLett.96.036801,PhysRevB.73.125411,PhysRevB.78.165411,PhysRevB.77.115109} without taking into account the
spin-orbit (SO) interaction.
It was found that the localized impurity
potential gives the resonant states in the spectrum of graphene.
They can be located near the Dirac point in the case of relatively
strong impurity potential, and this is quite unusual for the semiconductor physics.
In the case of carbon vacancy, there appears a local energy level at the Dirac point, with
$E=0$. In all of these works, the main attention has been paid to a finite density
of impurities and defects leading to modification of the density of
states in graphene.

In this paper we mostly concentrate on the problem of single impurity taking into account the
internal SO interaction. The SO interaction opens a
gap in the electron energy spectrum \cite{PhysRevLett.95.226801}. However, it was found that the magnitude of SO-induced gap
is very small in graphene \cite{PhysRevB.74.165310,PhysRevB.74.155426,PhysRevB.75.041401,PhysRevB.80.235431}, and therefore it would be very difficult to observe this gap experimentally. Nevertheless,
the problem exists: how this small gap would affect the behavior of impurity states
in the vicinity of Dirac point?

We demonstrate that the SO gap induces appearance of additional
impurity states corresponding to very weak impurity potential. As
a result, the SO gap in graphene can be experimentally
unobservable due to a large number of adsorbed light atoms at the
graphene surface creating the impurity states in the gap.

\section{Model}

We use the following Hamiltonian, which describes electrons near the Dirac point $K$, with the
intrinsic SO interaction \cite{PhysRevLett.95.226801}
\begin{eqnarray}
\label{1}
\hat{H}_{\bf k}=\left( \begin{array}{cc}
\sigma _z\Delta  & vk_- \\
vk_+ & -\sigma _z\Delta
\end{array} \right) ,
\end{eqnarray}
where $\Delta $ is the band splitting related to the SO interaction, $v$ is the
velocity parameter, and we denote $k_\pm =k_x\pm ik_y$. The matrix form of \eref{1} is due
to the choice of wavefunction basis corresponding to different sublattices
in the lattice of graphene. The basis functions of Hamiltonian \eref{1} are
\begin{eqnarray}
\label{2}
\left| {\bf k}(1,2)\sigma \right>
=\sum _{s\in (A,B)}\; e^{i({\bf k}_0+{\bf k})\cdot {\bf r}_i}\, \psi _s({\bf r })\, \left| \sigma \right> ,
\end{eqnarray}
where
$\psi _s({\bf r})$ is the tight-binding electron state at site $s$ belonging to
sublattice A or B, ${\bf k}_0$ is the wave vector corresponding to the chosen Dirac point,
and $\sigma =\uparrow ,\downarrow $ refers to spin up and down states, respectively.
The eigenvalues of Hamiltonian \eref{1} are
$E_{k1,2}=\pm \varepsilon _k$, where $\varepsilon _k= \left( \Delta ^2+v^2k^2\right) ^{1/2}$,
so that the value of $2|\Delta |$ is the energy gap.
The Hamiltonian for the other Dirac point $K'$ differs from \eref{1} by the opposite sign in the diagonal
terms. Thus, the results for the point $K'$ are can be found by using calculations with the
Hamiltonian \eref{1} and reverting the sign of $\Delta $.

Since the SO interaction in \Eref{1}
does not mix the spins, one can consider separately spin up and down channels.
For the spin up electrons the Hamiltonian can be presented as
it gives the Hamiltonian for up-spin  electrons
\begin{eqnarray}
\label{3}
\hat{H}_{{\bf k}\uparrow }
=\tau _z\Delta +v\bta \cdot {\bf k},
\end{eqnarray}
where $\tau _i$ are the Pauli matrices acting in the space of sublattices A and B.
For the down-spin Hamiltonian, the sign of $\Delta $ in \Eref{3} is opposite.
We consider first the spin up Hamiltonian \eref{3}.

\section{Nonmagnetic impurity}

Let us consider the impurity state in the case of a single impurity,
described by the perturbation localized in one of the sublattices. In the
continuous model under consideration it corresponds to the perturbation
at ${\bf r}=0$ in sublattice A
\begin{eqnarray}
\label{4}
\hat{V}_\uparrow ({\bf r})=\left( \begin{array}{cc}
V_0\, \delta ({\bf r})  & 0\\
0 & 0
\end{array} \right) .
\end{eqnarray}
The matrix of perturbation \eref{4} in the basis functions of Hamiltonian
$\hat{H}_{{\bf k}\uparrow }$ is
\begin{eqnarray}
\label{5}
\hat{V}_{{\bf kk'}\uparrow }
=\left( \begin{array}{cc}
V_0  & 0\\
0 & 0
\end{array} \right)
\equiv \hat{V}_{0\uparrow }.
\end{eqnarray}
The effect of perturbation $\hat{V}_{{\bf kk'}_\uparrow }$ on the energy spectrum in all
orders of magnitude
can be described using the $T$-matrix method \cite{Ziman}.
In the general case, the equation for the $T$-matrix is
\begin{equation}
\label{6}
\hat{T}_{\bf kk'}(\varepsilon )=\hat{V}_{\bf kk'}
+\sum _{{\bf k}_1}\hat{V}_{{\bf kk}_1}\, \hat{G}_{{\bf k}_1}(\varepsilon )\,
\hat{T}_{{\bf k}_1{\bf k'}}(\varepsilon ) ,
\end{equation}
where $\hat{G}_{\bf k}(\varepsilon )=\left( \varepsilon -\hat{H}_{\bf k}\right) ^{-1}$ is
the Green's function of Hamiltonian $\hat{H}_{\bf k}$.
Using \eref{3} we find the Green's function for spin up electrons
\begin{eqnarray}
\label{7}
\hat{G}_{{\bf k}\uparrow }(\varepsilon )
=\frac{\varepsilon +\tau _z\Delta +v\bta \cdot {\bf k}}{\varepsilon ^2-\varepsilon _k^2}\, .
\end{eqnarray}
In the following we assume that the energy parameter includes a small imaginary part,
$\varepsilon \to \varepsilon +i\delta \, {\rm sgn}\, \varepsilon $,
which corresponds to the choice of retarded Green's function.
Using \eref{5} and \eref{7} we find
\begin{eqnarray}
\label{8}
\hat{T}_\uparrow (\varepsilon )
=\left[ 1-\hat{V}_{0\uparrow }\sum _{\bf k}\hat{G}_{{\bf k}\uparrow }(\varepsilon )\right] ^{-1}
\hat{V}_{0\uparrow }
\equiv \left[ \hat{1}-\hat{V}_{0\uparrow }\,
\hat{F}_\uparrow (\varepsilon )\right] ^{-1}\hat{V}_{0\uparrow },
\hskip0.2cm
\end{eqnarray}
and we have to calculate
\begin{eqnarray}
\label{9}
\hat{F}_\uparrow (\varepsilon )\equiv \sum _{\bf k}\hat{G}_{{\bf k}\uparrow }(\varepsilon )
=-\frac{\varepsilon +\tau _z\Delta }{4\pi v^2}\;
\ln \frac{v^2k_m^2+\Delta ^2-\varepsilon ^2}{\Delta ^2-\varepsilon ^2}
\nonumber \\
\simeq -\frac{\varepsilon +i\delta \, {\rm sgn }\, \varepsilon +\tau _z\Delta }{4\pi v^2}\;
\ln \frac{v^2k_m^2+\Delta ^2-\varepsilon ^2 -2i|\varepsilon |\delta }
{\Delta ^2-\varepsilon ^2-2i|\varepsilon |\delta }\; ,
\end{eqnarray}
where the upper limit (cutoff) of integration over momentum $k_m$ is introduced.
It corresponds to the region of linearity of the spectrum near the Dirac point.
Assuming $v^2k_m^2\gg |\Delta ^2-\varepsilon ^2|$ we obtain
\begin{eqnarray}
\label{10}
\hat{F}_\uparrow (\varepsilon )
\simeq -\frac{\varepsilon +i\delta \, {\rm sgn }\, \varepsilon +\tau _z\Delta }{4\pi v^2}
\left[ \ln \frac{v^2k_m^2}{\left[ \left( \Delta ^2-\varepsilon ^2\right) ^2+4\varepsilon ^2\delta ^2\right] ^{1/2}}
+i(\varphi_1-\varphi _2)\right] ,\hskip0.5cm
\end{eqnarray}
where $\varphi _{1,2}$ are the angles related to the phase of complex function $\hat{F}(\varepsilon )$,
which can be made analytical in the whole complex plane of $\varepsilon $ after proper choice of
cuts in this plane.
We assume that te cut is made along the real axis from $-|\Delta |$ to
$+|\Delta |$. Then the phases can be chosen
\begin{eqnarray}
\label{11}
&&\varphi _1=-\tan ^{-1}\frac{2|\varepsilon |\delta }{v^2k_m^2}\, ,
\\
&&\varphi _2=
\left\{ \begin{array}{cc}
\pi+\tan ^{-1}\displaystyle{\frac{2|\varepsilon |\delta }{|\varepsilon ^2-\Delta ^2|}}, & \varepsilon ^2>\Delta ^2, \\
\\
-\tan ^{-1}\displaystyle{\frac{2|\varepsilon |\delta }{|\varepsilon ^2-\Delta ^2|}}, & \varepsilon ^2<\Delta ^2.
\end{array}\right.
\end{eqnarray}
The real part of \Eref{9}
\begin{eqnarray}
\label{13}
{\rm Re}\, \hat{F}_\uparrow (\varepsilon )
\simeq -\frac{\varepsilon +\tau _z\Delta }{4\pi v^2}\;
\ln \frac{v^2k_m^2}{\left[\left( \Delta ^2-\varepsilon ^2\right) ^2+4\varepsilon ^2\delta ^2\right] ^{1/2}}\; .
\end{eqnarray}
In the limit of $\delta \to 0$ (corresponding to small impurity density)
\begin{eqnarray}
\label{14}
{\rm Re}\, \hat{F}_\uparrow (\varepsilon )
\simeq -\frac{\varepsilon +\tau _z\Delta }{4\pi v^2}\;
\ln \frac{v^2k_m^2}{\left| \Delta ^2-\varepsilon ^2\right|}\, ,
\end{eqnarray}
\begin{eqnarray}
\label{15}
{\rm Im}\, \hat{F}_\uparrow (\varepsilon )\simeq
\left\{ \begin{array}{cc}
\displaystyle{\frac{\varepsilon +\tau _z\Delta }{4v^2}}, & \varepsilon ^2>\Delta ^2,\\
\\
0, & \varepsilon ^2<\Delta ^2,
\end{array} \right. .
\end{eqnarray}
The functions ${\rm Re}\, F_{11,\uparrow }(\varepsilon )$ and
${\rm Im}\, F_{11,\uparrow }(\varepsilon )$ are
presented in \fref{1} and \fref{2}.

\begin{figure}[h]
\begin{center}
\includegraphics[scale=0.4]{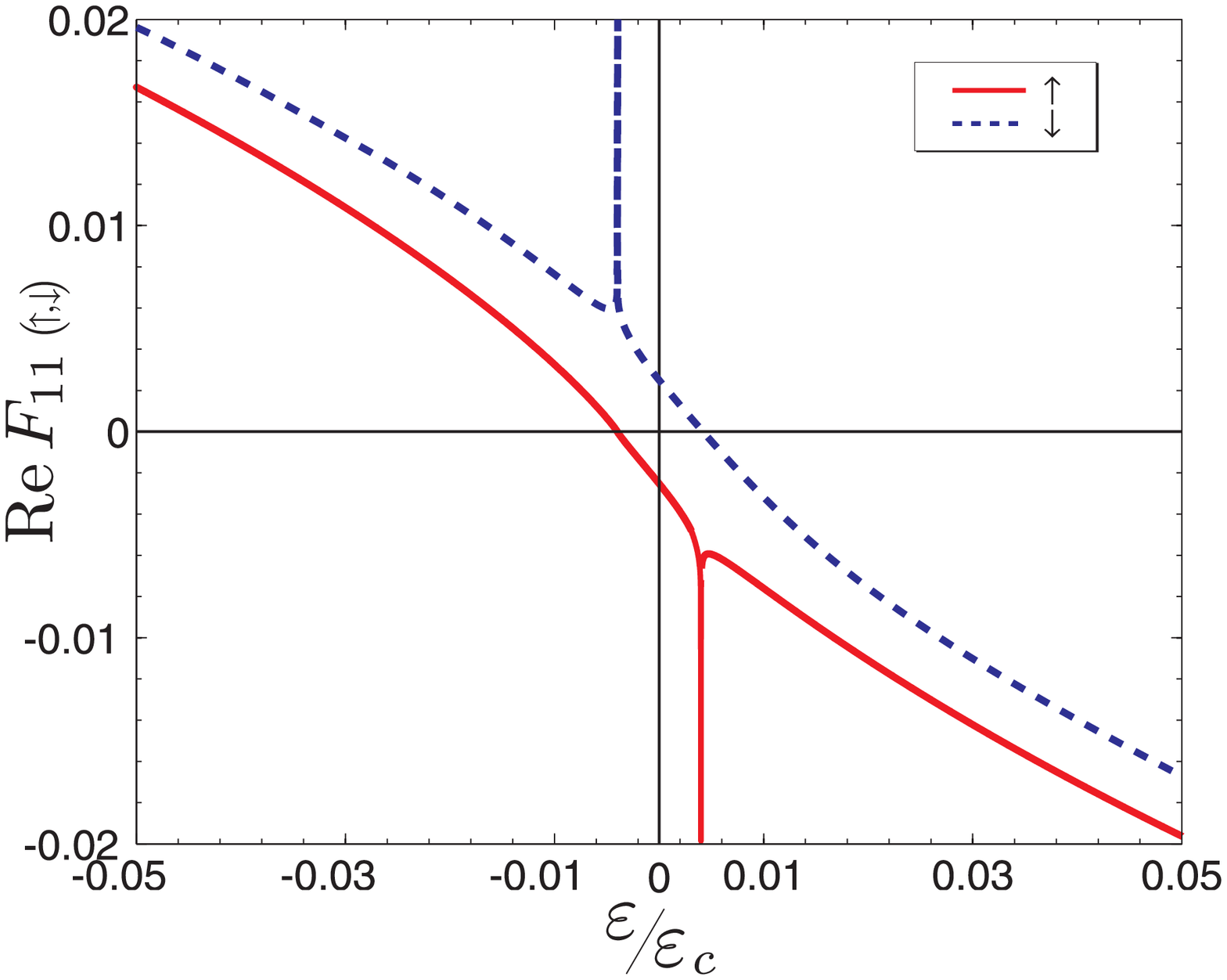}
\caption{The dependence ${\rm Re}\, F_{11}(\varepsilon )$ for spin-up and spin down states,
related to the Dirac point $K$ in the case of non-magnetic impurity.}
\label{fig1}
\end{center}
\end{figure}

\begin{figure}[h]
\begin{center}
\includegraphics[scale=0.4]{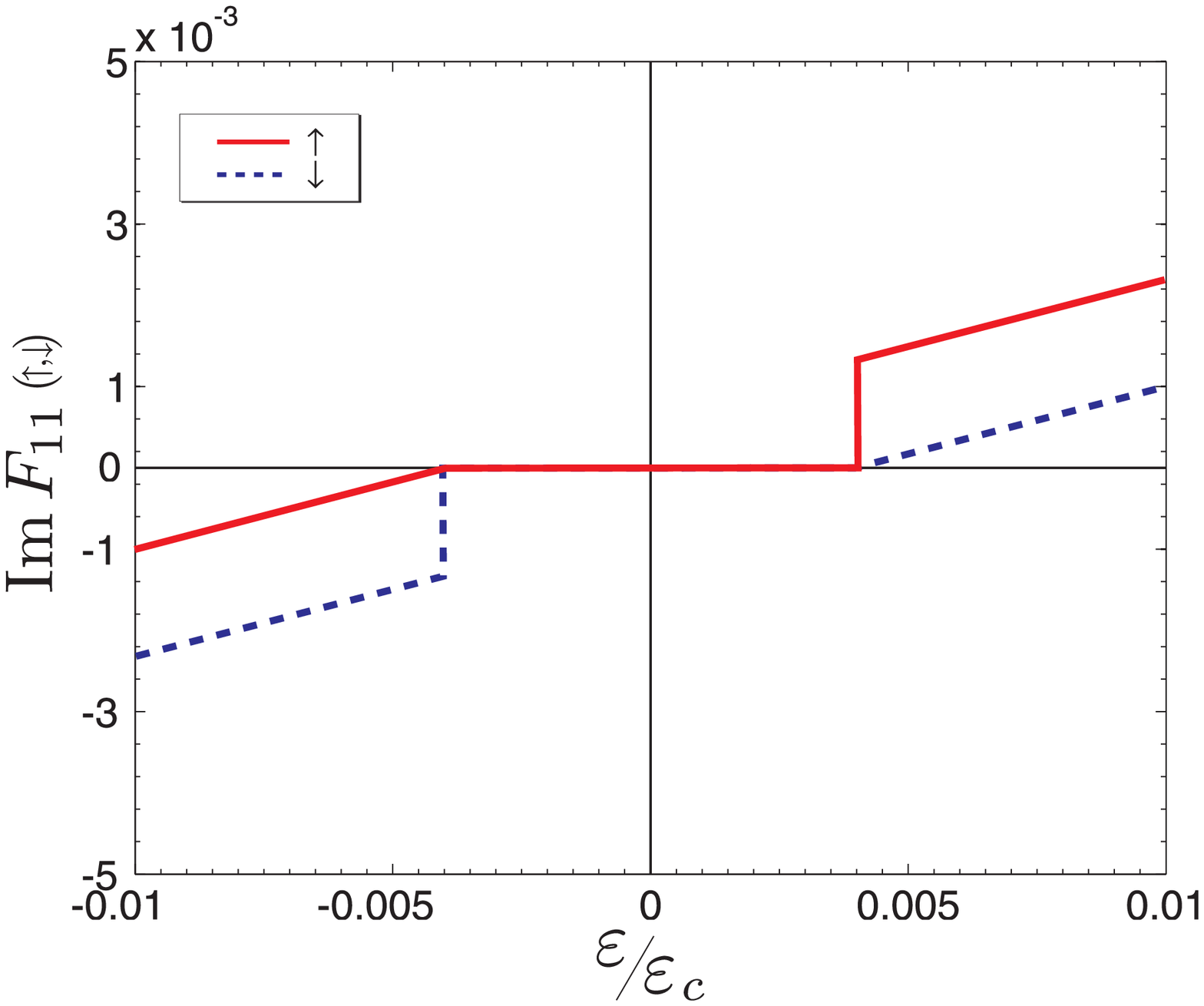}
\caption{The imaginary part ${\rm Im}\, F_{11}(\varepsilon )$ for spin-up and spin down states,
related to the point $K$ in the case of non-magnetic impurity.}
\label{fig2}
\end{center}
\end{figure}

The matrices $\hat{V}_{0\uparrow }$ and $\hat{F}_\uparrow (\varepsilon )$ are both diagonal.
Therefore, $\hat{T}(\varepsilon )$
calculated from \Eref{8} is diagonal, too
\begin{eqnarray}
\label{16}
\hat{T}_\uparrow (\varepsilon )={\rm diag} \left\{
\frac{V_{0\uparrow }}{1-V_{0\uparrow }\, {\rm Re}\, F_{11,\uparrow }(\varepsilon )
-iV_{0\uparrow }\, {\rm Im}\, F_{11,\uparrow }(\varepsilon )}\, ,\, 0\right\}\, .
\hskip0.2cm
\end{eqnarray}
The location of impurity level is determined by the pole of $T$-matrix
\begin{eqnarray}
\label{17}
1-V_{0\uparrow }\, {\rm Re}\, F_{11,\uparrow }(\varepsilon )=0.
\end{eqnarray}
The dependence ${\rm Re}\, F_{11,\uparrow }(\varepsilon )$ presented in \fref{1}
can be used for the graphical solution of \Eref{17}.
Using \eref{14} we find
\begin{eqnarray}
\label{18}
1+V_{0\uparrow }\, \frac{\varepsilon +\Delta }{4\pi v^2}\;
\ln \frac{v^2k_m^2}{\left| \Delta ^2-\varepsilon ^2\right|}=0.
\end{eqnarray}
Thus, the equation for the impurity level is
\begin{eqnarray}
\label{19}
\varepsilon _\uparrow
=-\Delta -\frac{4\pi v^2}{V_0\ln \displaystyle{\frac{v^2k_m^2}
{\left| \Delta ^2-\varepsilon _\uparrow ^2\right|}}}\, .
\end{eqnarray}
This equation has several solutions for the same potential $V_{0\uparrow }$.

The Hamiltonian describing the spin down states
\begin{eqnarray}
\label{20}
\hat{H}_{{\bf k}\downarrow }
=-\tau _z\Delta +v\bta \cdot {\bf k},
\end{eqnarray}
differs from \eref{2} only
by the sign of gap parameter $\Delta $.
In the case of nonmagnetic impurity the perturbation
$\hat{V}_\downarrow ({\bf r})=\hat{V}_\uparrow ({\bf r})$.
Performing the same calculations as before with the substitution $\Delta \to -\Delta $ we find
the equation for impurity level corresponding to the spin-down state
\begin{eqnarray}
\label{21}
\varepsilon _\downarrow =\Delta -\frac{4\pi v^2}{V_0\ln \displaystyle{ \frac{v^2k_m^2}
{\left| \Delta ^2-\varepsilon _\downarrow ^2\right|}}}\, .
\end{eqnarray}
The spin up and down states corresponding to solution of \Eref{19} and \Eref{21}, respectively,
are split in energy.
The magnitude of splitting is $2|\Delta |$,
which of course is very small as was discussed in the Introduction. Nevertheless, considering
the impurity states in the model with one Dirac point we come to a weakly magnetized
state at the nonmagnetic impurity. This nonequivalence of spin up and down states is exactly
compensated
by the states related to another Dirac point, $K'$, for which the Hamiltonians of spin up and down states
differ by the sign of $\Delta $ from those in \Eref{3} and \Eref{21} \cite{PhysRevLett.95.226801}.
Thus, the magnetization of the localized state is absent if we take into account
both nonequivalent Dirac points.

\begin{figure}[h]
\begin{center}
\includegraphics[scale=0.4]{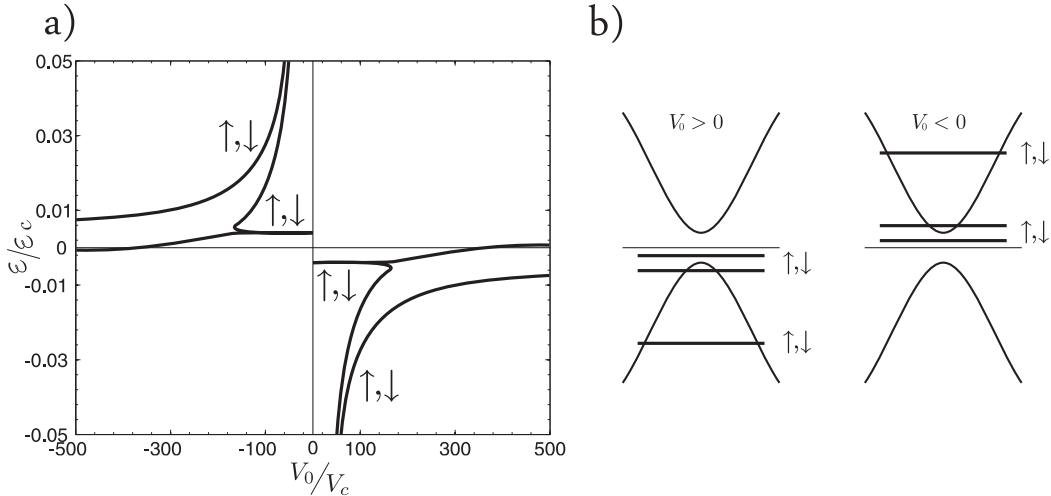}
\caption{The location of impurity levels as a function the impurity strength parameter $V_0$ (a)
and the schematic presentation of the impurity levels (b) in the case of non-magnetic impurity.}
\end{center}
\end{figure}

The numerical solutions of \Eref{19} and \Eref{21} describing the
states related to the Dirac point $K$, as well as the
corresponding solutions related to the Dirac point $K'$ are
presented in \Fref{3},a. The schematic representation of the
levels is presented in \Fref{3},b. All the levels are spin
degenerate due to the overlapping solutions of equations, related
to the nonequivalent Dirac points. We used the parameters with
much larger SO gap $\Delta $ to visualize better the character of
solutions. If the state is located within the gap, the impurity
level is discrete. For the level with energy $|\varepsilon
_\uparrow |>\Delta $, it is a resonant state of width ${\rm Im}\,
F_\uparrow (\varepsilon _\uparrow )$. It should be noted that
${\rm Im}\, F_\uparrow (\varepsilon )$ is not the density of
states in graphene because \Eref{9} does not include the trace
over sublattices.

It should be noted that both impurity states with energies
$\varepsilon _\uparrow $ and $\varepsilon _\downarrow $ are mostly
localized at the site of A sublattice in accordance with the
assumption that the impurity potential \eref{3} is located on the
A-site. However, the real spread of the wavefunction can be much
larger than the distance between nearest A and B sites.

\section{Magnetic impurity}

In the case of magnetic impurity we choose perturbation, which is different in
sign for the spin up and down electrons, $V_{{\bf kk'}\downarrow }=-V_{{\bf kk'}\uparrow }$,
and $V_{{\bf kk'}\uparrow }$ is described by \Eref{5}.
The resulting impurity levels related to the Dirac points $K$ and $K'$ are
to be found from the following equations
\begin{eqnarray}
\label{22}
\varepsilon ^{K}_{\uparrow ,\downarrow }
=\mp \Delta \mp \frac{4\pi v^2}{V_0\ln \displaystyle{ \frac{v^2k_m^2}
{\left| \Delta ^2-\varepsilon _{\uparrow ,\downarrow }^2\right|}}}\, .
\end{eqnarray}
\begin{eqnarray}
\label{23}
\varepsilon ^{K'}_{\uparrow ,\downarrow }
=\pm \Delta \mp \frac{4\pi v^2}{V_0\ln \displaystyle{\frac{v^2k_m^2}
{\left| \Delta ^2-\varepsilon _{\uparrow ,\downarrow }^2\right|}}}\, .
\end{eqnarray}
There is the spin splitting of the states related to the $K$ point,
which does not vanish as $\Delta \to 0$, and the resulting magnetization of
impurity states is not compensated by another point $K'$.
Corresponding numerical solutions of equations \eref{22} and \eref{23} are presented
in \Fref{4}.

\begin{figure}[h]
\begin{center}
\includegraphics[scale=0.4]{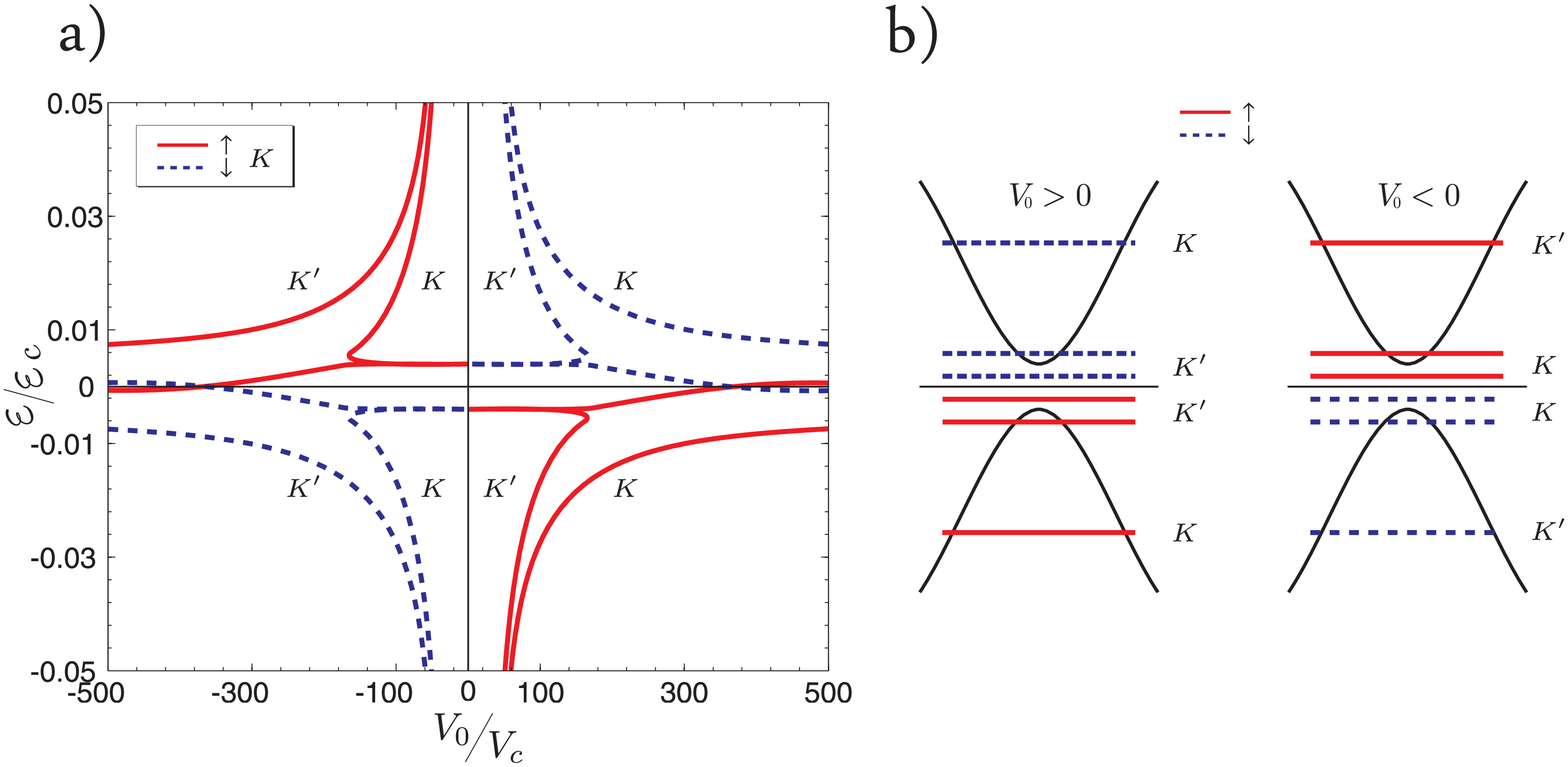}
\caption{The same dependences as in \Fref{3} in the case of magnetic impurity.}
\end{center}
\end{figure}

\section{Nonmagnetic impurity with strong SO interaction}

Using the same formalism, one can consider the impurity, which locally enhances
the internal SO interaction. In this case the corresponding perturbation for $K$ point
has the form
\begin{eqnarray}
\label{24}
\hat{V}_\Delta =\left( \begin{array}{cc}
\sigma _z\Delta _0 & 0 \\
0 & 0
\end{array} \right) ,
\end{eqnarray}
where $\Delta _0$ is the local
After substitution of $V_0\to \Delta _0$, the solution for the $K$ point does not differ
from the case of magnetic impurity considered above.
However, for the point $K'$ we have to change the sign of $\Delta _0$.
As a result we obtain
\begin{eqnarray}
\label{25}
\varepsilon ^{K}_{\uparrow ,\downarrow }
=\mp \Delta \mp \frac{4\pi v^2}{\Delta _0\ln \displaystyle{ \frac{v^2k_m^2}
{\left| \Delta ^2-\varepsilon _{\uparrow ,\downarrow }^2\right|}}}\, .
\end{eqnarray}
\begin{eqnarray}
\label{26}
\varepsilon ^{K'}_{\uparrow ,\downarrow }
=\pm \Delta \pm \frac{4\pi v^2}{\Delta _0\ln \displaystyle{ \frac{v^2k_m^2}
{\left| \Delta ^2-\varepsilon _{\uparrow ,\downarrow }^2\right|}}}\, .
\end{eqnarray}
In this case the spin splitting of the states related to the point $K$ does
not vanish as $\Delta \to 0$ but the resulting local magnetization is
absent due to the states from point $K'$. The solutions of equations \eref{25}
and \eref{26} are presented in \fref{5}.

\begin{figure}[h]
\begin{center}
\includegraphics[scale=0.4]{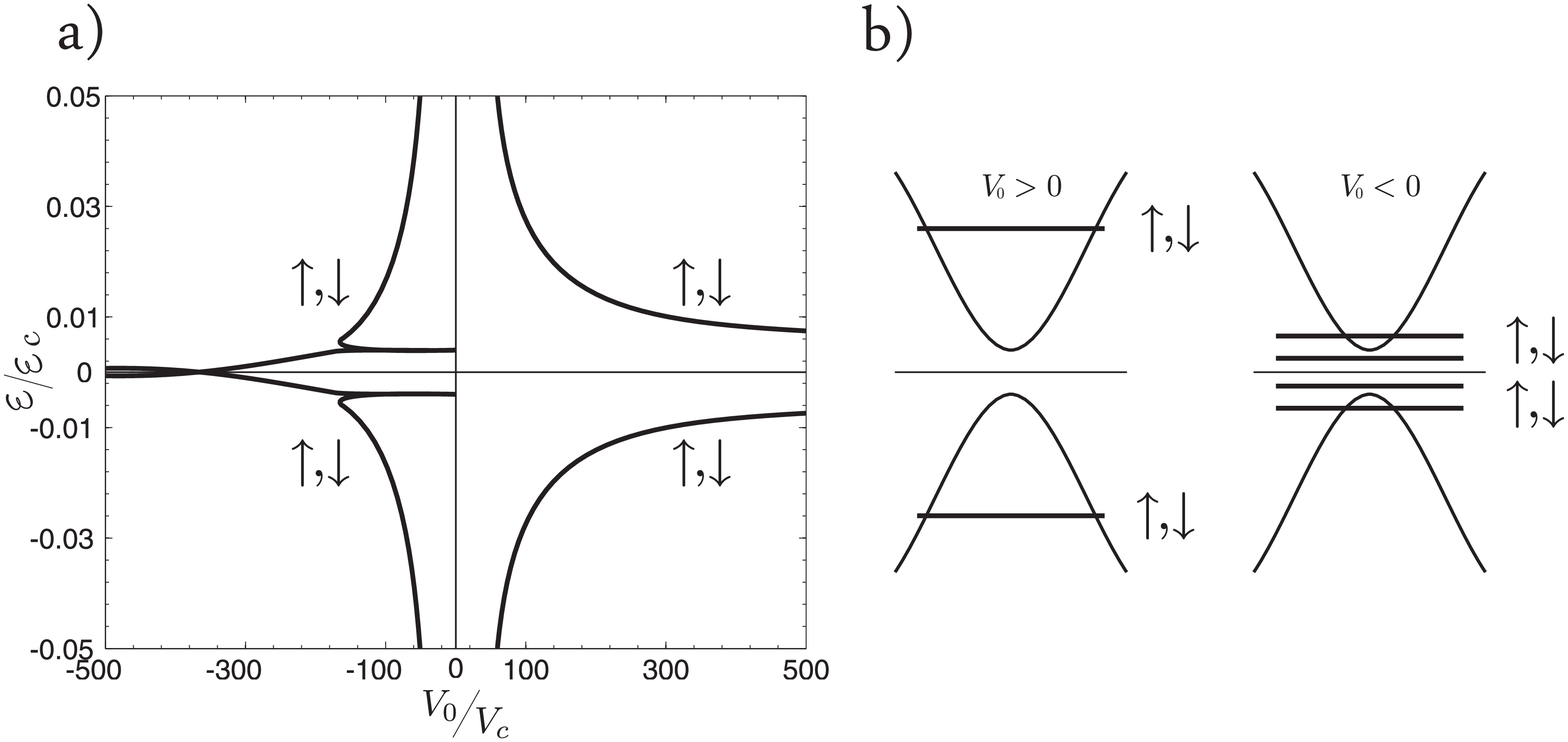}
\caption{The same dependences as in \Fref{3} in the case of spin-orbit impurity.}
\end{center}
\end{figure}

\section{Wave function of the localized impurity state}

Using the Schr\"odinger equation for the wave function, one can find
the equation for impurity state \cite{Ziman}
\begin{eqnarray}
\label{27}
\psi _i({\bf r})=\int d^2{\bf r'}\, \hat{G}({\bf r,r'};\varepsilon )\, \hat{V}({\bf r'})\, \psi _i({\bf r'}),
\end{eqnarray}
where the Green function of free electrons in coordinate representation obeys
\begin{eqnarray}
\label{28}
\left( \varepsilon -\hat{H}_{\bf r}\right) \hat{G}({\bf r,r'};\varepsilon ) =\delta ({\bf r}-{\bf r'}).
\end{eqnarray}
For the impurity located at ${\bf r}=0$ we obtain from \eref{27}
\begin{eqnarray}
\label{29}
\psi _i({\bf r})=\hat{G}({\bf r},0;\varepsilon )\, \hat{T}(\varepsilon)\, \psi _0(0),
\end{eqnarray}
where $\psi _0({\bf r})$ is the eigenfunction of unperturbed Hamiltonian \eref{1}
\begin{eqnarray}
\label{30}
\left( \varepsilon -\hat{H}_{\bf r}\right) \, \psi _0({\bf r})=0.
\end{eqnarray}

The Green function for Hamiltonian \eref{3} has been calculated before \cite{PhysRevB.74.224438}
\begin{eqnarray}
\label{31}
\hat{G}_{\uparrow }({\bf r,r'};\varepsilon )
=-\frac{i(\varepsilon -\Delta \tau _z)}{4v^2}
H_0^{(1)}\left( \frac{|{\bf r}-{\bf r'}| \sqrt{\varepsilon ^2-\Delta ^2}}{v}\right)
\nonumber \\
+({\bf r}-{\bf r'})\cdot \bta \;
\frac{\sqrt{\varepsilon ^2-\Delta ^2}}{4v^2|{\bf r}-{\bf r'}|}
H_1^{(1)}\left( \frac{|{\bf r}-{\bf r'}|\sqrt{\varepsilon ^2-\Delta ^2}}{v}\right) .
\end{eqnarray}
Thus, for the spin-up impurity state with energy $\varepsilon _\uparrow $
\begin{eqnarray}
\label{32}
|\psi _{i\uparrow }({\bf r})| ^2\sim
\left| H_0^{(1)}\left( \frac{r\, \sqrt{\varepsilon _\uparrow ^2-\Delta ^2}}{v}\right) \right| ^2
+\left| H_1^{(1)}\left( \frac{r\, \sqrt{\varepsilon _\uparrow ^2-\Delta ^2}}{v}\right) \right| ^2
\sim e^{-2r/R_0},\hskip0.5cm
\end{eqnarray}
where $H_\nu ^{(1)}(z)$ are the Hankel functions \cite{Abramowitz} and
$R_{0\uparrow }=v/\left| \varepsilon _\uparrow ^2-\Delta ^2\right| ^{1/2}$ is the characteristic radius of the impurity
wavefunction.

The function $\psi _{0}({\bf r})$ in \Eref{29} can be calculated using \eref{30}. Denoting the bispinor
components $\psi _0^T({\bf r})=\left( \varphi ^T({\bf r}),\; \chi ^T({\bf r})\, \right) $
we find the relation
\begin{eqnarray}
\label{33}
\chi ({\bf r})=\frac{-iv(\partial _x+i\partial _y)}{\Delta +\varepsilon }\; \varphi ({\bf r}).
\end{eqnarray}
The equation for the function $\varphi _\uparrow $ with the Hamiltonian \eref{3} in polar coordinates
$(r,\alpha )$ reads
\begin{eqnarray}
\label{34}
\left( \frac{\partial ^2}{\partial r^2}+\frac1{r}\, \frac{\partial}{\partial r}+\frac1{r^2}\,
\frac{\partial ^2}{\partial \alpha ^2}-\kappa _\uparrow ^2\right)
\varphi _\uparrow (r,\alpha )=0,
\end{eqnarray}
which has the following solutions decaying at large $r$
\begin{eqnarray}
\label{35}
\varphi _\uparrow (r,\alpha )=K_m(\kappa _\uparrow r)\, e^{im\alpha },
\end{eqnarray}
where $\kappa _\uparrow =\left( \Delta ^2-\varepsilon _\uparrow ^2\right) /v^2$,
$K_m(z)$ is the modified Bessel function and $m\in Z$.
Then the corresponding $\chi _\uparrow $-component can be found from \eref{33} and \eref{35}
\begin{eqnarray}
\label{36}
\chi _\uparrow (r, \alpha )=-\frac{iv\kappa _\uparrow }{\Delta +\varepsilon _\uparrow }
K_{m+1}(\kappa _\uparrow r)\, e^{i(m+1)\alpha }.
\end{eqnarray}
Thus, up to the normalization we have
\begin{eqnarray}
\label{37}
\psi _{0\uparrow m}(r,\alpha )=\left( \begin{array}{c}
(\Delta +\varepsilon _\uparrow )\, K_m(\kappa _\uparrow r)\, e^{im\alpha }
\\
-iv\kappa _\uparrow \, K_{m+1}(\kappa _\uparrow r)\, e^{i(m+1)\alpha }
\end{array}\right)
\end{eqnarray}
It should be noted that for $\psi _0(0)$ in \Eref{29} we have to use the cutoff at small
distance $r_c\sim k_{max}^{-1}\gg a_0/\pi $, where $a_0$ is the lattice constant and
$k_{max}$ is the upper limit for the Dirac model in graphene.
It means that using the Dirac model we can find the impurity wave function at distances
much larger than $a_0$. On the other hand, as we see from \eref{32}, the radius of the
impurity states $R_0$ near the Dirac point is much larger than this limit.

\section{Conclusions}

We have calculated the energies and wave functions of impurity
states near the Dirac points in graphene taking into account the
SO interaction. The calculations show the SO-induced spin
splitting of these states. The existence of two nonequivalent
Dirac points in the Brillouin zone leads to the spin degeneracy of
the states with different spin. It should be noted that in
principle the valley degeneracy can be broken by inhomogeneous
deformations, which would result in the appearance of local
magnetization.

\ack
This work was supported by the Polish Ministry of Science and Higher Education
as a research project in years 2007 -- 2010.

\section*{References}


\begin{thebibliography}{10}


\bibitem{K.S.Novoselov10222004}
Novoselov K~S, Geim A~K, Morozov S~V, Jiang D, Zhang Y, Dubonos S~V, Grigorieva
  I~V and Firsov A~A 2004 {\em Science\/} {\bf 306} 666--669

\bibitem{NovoselovNature05}
Novoselov K~S, Geim A~K, Morozov S~V, Jiang D, Katsnelson M~I, Grigorieva I~V,
  Dubonos S~V and Firsov A~A 2005 {\em Nature\/} {\bf 438} 197--200

\bibitem{GeimNatMater07}
Geim A~K and Novoselov K~S 2007 {\em Nat Mater\/} {\bf 6} 183--191

\bibitem{Katsnelson200720}
Katsnelson M~I 2007 {\em Materials Today\/} {\bf 10} 20--27 ISSN 1369-7021

\bibitem{SchedinNature07}
Schedin F, Geim A~K, Morozov S~V, Hill E~W, Blake P, Katsnelson M~I and
  Novoselov K~S 2007 {\em Nat Mater\/} {\bf 6} 652--655

\bibitem{RevModPhys.81.109}
Castro~Neto A~H, Guinea F, Peres N~M~R, Novoselov K~S and Geim A~K 2009 {\em
  Rev. Mod. Phys.\/} {\bf 81} 109--162

\bibitem{PhysRevLett.96.036801}
Pereira V~M, Guinea F, Lopes~dos Santos J~M~B, Peres N~M~R and Castro~Neto A~H
  2006 {\em Phys. Rev. Lett.\/} {\bf 96} 036801

\bibitem{PhysRevB.73.125411}
Peres N~M~R, Guinea F and Castro~Neto A~H 2006 {\em Phys. Rev. B\/} {\bf 73}
  125411

\bibitem{PhysRevB.78.165411}
Hu B~Y~K, Hwang E~H and Das~Sarma S 2008 {\em Phys. Rev. B\/} {\bf 78} 165411

\bibitem{PhysRevB.77.115109}
Pereira V~M, Lopes~dos Santos J~M~B and Castro~Neto A~H 2008 {\em Phys. Rev.
  B\/} {\bf 77} 115109

\bibitem{PhysRevLett.95.226801}
Kane C~L and Mele E~J 2005 {\em Phys. Rev. Lett.\/} {\bf 95} 226801

\bibitem{PhysRevB.74.165310}
Min H, Hill J~E, Sinitsyn N~A, Sahu B~R, Kleinman L and MacDonald A~H 2006 {\em
  Phys. Rev. B\/} {\bf 74} 165310

\bibitem{PhysRevB.74.155426}
Huertas-Hernando D, Guinea F and Brataas A 2006 {\em Phys. Rev. B\/} {\bf 74}
  155426

\bibitem{PhysRevB.75.041401}
Yao Y, Ye F, Qi X~L, Zhang S~C and Fang Z 2007 {\em Phys. Rev. B\/} {\bf 75}
  041401

\bibitem{PhysRevB.80.235431}
Gmitra M, Konschuh S, Ertler C, Ambrosch-Draxl C and Fabian J 2009 {\em Phys.
  Rev. B\/} {\bf 80} 235431

\bibitem{Ziman}
Ziman J~M 1969 {\em Elements of Advanced Quantum Theory\/} (Cambridge Univ.
  Press, Cambridge)

\bibitem{PhysRevB.74.224438}
Dugaev V~K, Litvinov V~I and Barnas J 2006 {\em Phys. Rev. B\/} {\bf 74} 224438

\bibitem{Abramowitz}
Abramowitz M and Stegun I~A 1964 {\em Handbook of Mathematical Functions\/}
  Natl. Bur. Stand. Appl. Math. Ser. 55 (National Bureau of Standards,
  Washington, DC, 1964)
  
  \end{thebibliography}

\end{document}